\documentclass[aps,prb,reprint,showpacs]{revtex4-1}
\usepackage[dvipdfm]{graphicx}

\usepackage{amsmath}

\usepackage{bm}
\def\vb#1{{\bm#1}}
\def\v#1{\mathbf{#1}}			

\def\vv{\v{v}}

\def\va{\v{a}}

\def\r{\v{r}} 					% def. of vector "r"
\def\p{\v{p}} 					% def. of vector "p"
\def\k{\v{k}} 					% def. of vector "k"
\def\A{\v{A}}

\def\vOmega{\vb{\Omega}}

\def\vsigma{\vb{\sigma}}

\def\B{\v{B}}
\def\E{\v{E}}
\def\vv{\v{v}}

\def\va{\v{a}}

\def\div{{\rm div \,}}

\def\tpsi{\tilde{\psi}}
\def\tE{\tilde{E}}

\def\del{\partial}

\def\CB{\mathcal{B}}
\def\CE{\mathcal{E}}

\def\CA{\mathcal{A}}
\def\CF{\mathcal{F}}

\def\CL{\mathcal{L}}
\def\CJ{\mathcal{J}}

\def\ket#1{\left| #1 \right\rangle }
\def\bra#1{\left\langle  #1 \right| }

\begin{document}

%\preprint{}

\title{Renormalization of spin-rotation coupling}



\author{Mamoru Matsuo,
Jun'ichi Ieda, 
and Sadamichi Maekawa }
\affiliation{
Advanced Science Research Center, Japan Atomic Energy Agency, Tokai 319-1195, Japan \\
CREST, Japan Science and Technology Agency, Sanbancho, Tokyo 102-0075, Japan
}

\date{\today}

\begin{abstract}
We predict the enhancement of the spin-rotation coupling due to the interband mixing. 
The Bloch wavefunctions in the presence of mechanical rotation are constructed with the generalized crystal momentum
which includes a gauge potential arising from the rotation. 
Using the eight-band Kane model, the renormalized spin-rotation coupling is explicitly obtained. 
As a result of the renormalization, the rotational Doppler shift in electron spin resonance and the mechanical torque on an electron spin 
will be strongly modulated.% in materials with the large $g$ factor. 
\end{abstract}

%*72.25.-b	Spin polarized transport
% 72.25.Dc Spin polarized transport in semiconductors
%*85.75.-d Magnetoelectronics; spintronics: devices exploiting spin polarized transport or integrated magnetic fields
%*71.70.Ej Spin-orbit coupling, Zeeman and Stark splitting, Jahn-Teller effect
%*62.25.-g Mechanical properties of nanoscale systems
%
%
%

\maketitle %\maketitle must follow title, authors, abstract and \pacs

\section{Introduction}
A variety of coupling of an electron spin with other degrees is enhanced in solids.
The modulation of the electron $g$ factor in the Zeeman interaction due to the interband mixing has been widely studied in semiconductors\cite{Winkler2006}. Spin manipulation by $g$ factor engineering\cite{Salis2001,Kato2003} has attracted interest in the context of quantum information processing.
Spin-orbit interaction (SOI), the coupling of electron spin and orbital motion, is enhanced depending on the band structure.  
Enhancement of SOI is responsible for the spin Hall effect which plays an important role in the conversion between charge and spin current\cite{MaekawaEd2012} in the field of spintronics\cite{Zutic2004,MaekawaEd2006}.

An electron spin also couples to mechanical rotation. In the presence of mechanical rotation with frequency $\vb{\Omega}$, the spin angular momentum $\frac{\hbar}{2}\vb{\sigma}$ couples to the rotation as
$H_{S} = - \frac{\hbar}{2}\vb{\sigma}\cdot \vb{\Omega}$, which is known as the spin-rotation coupling\cite{Oliveira1962,Mashhoon1988,Hehl1990}.
The quantum mechanical nature of the Barnett and Einstein-de Haas effects \cite{Barnett1915,Einstein-deHaas1915} can be explained on the basis of the coupling\cite{Frohlich1993}.
The bare coupling of a spin and rotation is of great importance in neutron interferometry\cite{Rauch2000}
as well as in tests of general relativity using spin precession in gravitational field of a rotating body\cite{L-T,deSitter1916,Bertotti1987,Schiff1960,Everitt2011}.

%Recently, the interplay between spin and mechanical rotation has drawn much attention in spintronics.
Rapid progress in nanotechnology has allowed us to study the coupling of mechanical motion and nano-magnetic systems.
Einstein-de Haas effect in a NiFe film on a microcantilever\cite{Wallis2006} and mechanical torque due to spin flip on a torsion oscillator\cite{Zolfagharkhani2008} were observed in experiment. 
Theoretically, 
rotational doppler shift in magnetic resonance\cite{Lendinez2010} 
and effects of mechanical torque in nanostructure\cite{Mohanty2004,Kovalev030507,Calero2005,Bretzel2009,Jaafar2009prb,Chudnovsky2010,Bauer2010,Jaafar2009prl,Kovalev2011}
are studied. 
These phenomena essentially rely on the spin-rotation coupling, $H_{S}$. 
However, 
the renormalization of the coupling arising from the electronic structures
has not been considered so far.

In this paper, we theoretically investigate the renormalization of inertial effects on spin in a solid using the $\k \cdot \p$ perturbation with the generalized crystal momentum due to mechanical rotation. 
It is shown that the mechanically induced SOI, Darwin term and spin-rotation coupling are enhanced by the interband mixing of the conduction band and valence band states. 
The renormalized spin-rotation coupling is responsible for the enhancement of both
the frequency shift in electron spin resonance (ESR) 
and
the mechanical torque in spin precession. 

The outline of the paper is the following. 
In Sec. II, we formulate Bloch's theorem in the presence of mechanical rotation. 
In Sec. III, the envelop function approximation in a rotating frame is presented. 
In Sec. IV, we show that the renormalization of spin-rotation coupling due to the interband mixing by using the eight band Kane model. 
In Sec. V and VI, we discuss that the renormalization provides the enhancement of the frequency shift in ESR and the mechanical torque in spin precession. 
The order of the introduction of the gauge field and the projection to the conduction electron is discussed in Sec. VII. 
The paper ends with a few concluding remarks in Sec. VIII. 
$SU(2) \times U(1)$ gauge theory in rigidly accelerated frames are summarized in Appendix. 

\section{Bloch's theorem in a rotating frame}
Let us consider an electron in a periodic potential in the presence of mechanical rotation. 
We start with a simple Hamiltonian in an inertial frame given by 
\begin{eqnarray}
H_{0}= \frac{\p_{0}^{2}}{2m}+V_{0}(\r_{0}),
\end{eqnarray}
where $V_{0}(\r_{0})$ is a microscopic periodic crystal potential, $\r_{0}$ and $\p_{0}= i\hbar \frac{\del}{\del \r_{0}}$ are the coordinate and momentum operator in the inertial frame, respectively.
Performing the unitary transformation:
\begin{eqnarray}
U=\exp[i \v{J}\cdot \vOmega t/\hbar],
\end{eqnarray}
 with $\v{J} = \r_{0} \times \p_{0} + \frac{\hbar}{2}\vsigma$ being the generator of the rotation,
one obtains the Hamiltonian in a rotating frame\cite{Frohlich1993}:
\begin{eqnarray}
H_{R} = U H_{0} U^{\dagger} -i\hbar  U \frac{\del U^{\dagger}}{\del t} =H_{0} - \v{J} \cdot \vOmega ,
\end{eqnarray}
which reads 
\begin{eqnarray}
H_{R} = \frac{(\p - q \A_{g})^{2}}{2m} + V_{0}(\r) + q\phi_{g} + \mu_{B} \vsigma \cdot \frac{\B_{g}}2 \label{HR}
\end{eqnarray}
where $q=-e$ is the electron charge, $\mu_{B}= e\hbar/2m$ is the Bohr magneton, $\r$ and $\p$ are the coordinate and momentum operators in the rotating frame, respectively, 
$\phi_{g} = - \gamma_{0}^{-1} (\vOmega \times \r)^{2}/2$, and $\A_{g} = \gamma_{0}^{-1} \vOmega \times \r$ where $\gamma_{0} = -e/m$ is the electron charge-to-mass quotient.
The gravitomagnetic field $\B_{g}$ is defined by\cite{RyderGR2009} $\B_{g} = \nabla \times \A_{g} = 2 \gamma_{0}^{-1} \vOmega$.  
The gauge fields $\phi_{g}$ and $\A_{g}$ originate from $- \v{L} \cdot \vOmega=- (\r \times \p) \cdot \vOmega$ 
while the effective Zeeman interaction $\mu_{B} \vsigma \cdot \B_{g}/2$ originates from the spin-rotation coupling $H_{S}$.

Next, we consider Bloch's theorem in a rotating frame. 
When the rotation $\vOmega$ is applied, the crystal momentum is defined as 
\begin{eqnarray}
\hbar \k_{g} = \p -q\A_{g}
\end{eqnarray}
and the eigenfunctions in a periodic potential 
can be expressed as $\psi_{n,\k_{g}}(\r)=u_{n,\k_{g}}(\r) e^{i\k_{g} \cdot \r}$, where $u_{n,\k_{g}}(\r)$ has the periodicity of the crystal and $n$ is the band index. 
The above argument is verified by the magnetic translation group theory\cite{Zak1964}.
In the presence of a magnetic field $\B=\nabla \times \A$, the Hamiltonian is invariant under the magnetic translation group, 
and hence, the momentum is given by $\hbar \k = \p -q\A$ whose component does not commute with each other. 
Likewise, the Hamiltonian $H_{R}$ is invariant under the ``magnetic'' translation group in which the gauge potential $\A$ is replaced by $\A_{g}$.

\section{The envelope function approximation}
Once the Bloch eigenfunctions are obtained, the conventional $\k \cdot \p$ method can be generalized in the presence of the mechanical rotation. 
Here we add the slowly varying potential $V(\r)$ and the SOI due to the crystal potential $H_{\rm SO}$ to $H_{R}$:
\begin{eqnarray}
H_{R}' = H_{R} +V(\r)+H_{\rm SO},
\end{eqnarray}
where 
\begin{eqnarray}
H_{\rm SO} = \frac{\lambda}{\hbar} (\p - q \A_{g}) \cdot \vsigma \times (\nabla V_{0}). 
\end{eqnarray} 
The wave function $\Psi(\r)$ is expanded in terms of band-edge Bloch functions $u_{n',0}$ times spin eigenstates $\ket{\sigma'}$: 
\begin{eqnarray}
\Psi(\r)= \sum_{n'\sigma'} \psi_{n'\sigma'}(\r)u_{n'\v{0}}(\r) \ket{\sigma'},
\end{eqnarray}
where $ \psi_{n'\sigma'}(\r)$ is the slowly varying envelope function that modulates the quickly oscillating Bloch wave functions $u_{n'\v{0}}(\r)$. 
Multiplied from the left by $\bra{\sigma}u_{n\v{0}}^{*}$,  the equation 
\begin{eqnarray}
H_{R}' \Psi(\r) = E\Psi(\r)
\end{eqnarray}
can be reduced to the coupled equation of the envelop function: 
\begin{eqnarray}
\sum_{n'\sigma'}H^{{\rm EFA}}_{n'\sigma'} \psi_{n'\sigma'}(\r) = E \psi_{n\sigma}(\r)
\end{eqnarray}
with the multiband Hamiltonian:
\begin{eqnarray}
H^{\rm EFA}_{n'\sigma'} &=&  \left[E_{n'}(\v{0}) + \frac{\hbar^{2} \k_{g}^{2}}{2m} + V(\r)   \right]\delta_{nn'}\delta_{\sigma\sigma'} \nonumber\\
 &&+ \frac{\hbar \k_{g}}{m}\cdot \v{P}^{nn'}_{\sigma\sigma'} +  \Delta^{nn'}_{\sigma\sigma'} + \mu_{B}\vsigma \cdot \frac{\B_{g}}{2} \delta_{nn'},
\end{eqnarray}
where 
\begin{eqnarray}
\v{P}^{nn'}_{\sigma\sigma'} = \bra{n\sigma} [\p + \frac{\lambda}{\hbar}\vsigma \times \nabla V_{0}] \ket{n'\sigma'} \label{Pnn}
\end{eqnarray}
and 
\begin{eqnarray}
\Delta^{nn'}_{\sigma\sigma'}=\bra{n\sigma} \p \cdot \frac{\lambda}{\hbar}\vsigma \times \nabla V_{0} \ket{n'\sigma'}. \label{Dnn}
\end{eqnarray}
Here we assume $V(\r), \A_{g},$ and $\psi_{n'\sigma'}(\r)$ are slowly varying within one unit cell, and then they are taken out of the integral as constant. 
Using quasi-degenerate perturbation theory\cite{Winkler2006}, the infinite-dimensional Hamiltonian $H^{\rm EFA}_{n'\sigma'}$ can be converted into a finite-dimensional one in which the SOI and the effective Zeeman interaction are renormalized by the interband mixing. 
Thus, the mechanism of the renormalization of the spin-rotation coupling is quite similar to that of the conventional Zeeman interaction.

\section{$k\cdot p$ perturbation in a rotating body}
In the following, we consider the renormalization due to the interband mixing in zincblende-type semiconductors close to the Gamma point of the Brillouin zone using the $8\times 8$ Kane model 
which includes the $\k \cdot \p$ coupling between the $\Gamma^{c}_{6}$ conduction band and the $\Gamma^{v}_{8}$ and $\Gamma^{v}_{7}$ valence bands\cite{Winkler2006,Kane1957}: 
\begin{eqnarray}
&&H'_{8\times8} =
\left(\begin{array}{ccc}
H'_{6c6c} & H'_{6c8v} & H'_{6c7v} \\
H'_{8v6c} & H'_{8v8v} & H'_{8v7v} \\
H'_{7v6c} & H'_{7v8v} & H'_{7v7v} 
\end{array}\right) \nonumber \\
&&=
\left(\begin{array}{ccc}
(E_c + V) I_{2} & \sqrt3 P \v{T} \cdot \k_{g} & -\frac{P}{\sqrt3}  \vsigma \cdot \k_{g} \\
\sqrt3 P \v{T}^{\dagger} \cdot \k_{g} & (E_v + V) I_{4} & 0 \\
-\frac{P}{\sqrt3} \vsigma \cdot \k_{g} & 0 & (E_v - \Delta_0 + V) I_{2}
\end{array}\right), \nonumber \\ \label{8x8}
\end{eqnarray}
where $V=V(\r)$, $E_{c}$ and $E_{v}$ are the conduction and valence band edges, respectively.
From Eq. (\ref{Pnn}) and (\ref{Dnn}) we define the Kane momentum matrix element as 
$P=(\hbar/m)\bra{S}p_{x}\ket{X}$, and also define the spin-orbit gap as $\Delta_{0} = -(3i\hbar/4m^{2}c^{2})\bra{X}[\nabla V_{0} \times \p]_{y} \ket{Z}$, where $\ket{S}$ is the $s$-like conduction band state and $\ket{X}$ and $\ket{Z}$ are the $p$-like valence band states\cite{Winkler2006}.
Note that the conventional crystal momentum used in the $\k \cdot \p$ perturbation $\hbar \k= \p -q \A$ is replaced by $\hbar \k_{g}=\p -q \A_{g}$, which includes the inertial effect due to the gravitomagnetic field $\B_{g}$.
The matrices $\v{T}=(T_{x},T_{y},T_{z})$ are given by
\begin{eqnarray}
T_{x} &=&\frac1{3\sqrt2}
\left(\begin{array}{cccc}
-\sqrt3 & 0 & 1 & 0 \\
0 & -1 & 0 & \sqrt3
\end{array}\right), \\
T_{y} &=&\frac{-i}{3\sqrt2}
\left(\begin{array}{cccc}
\sqrt3 & 0 & 1 & 0 \\
0 & 1 & 0 & \sqrt3
\end{array}\right), \\
T_{z} &=&\frac{\sqrt2}{3}
\left(\begin{array}{cccc}
0 & 1 & 0 & 0 \\
0 & 0 & 1 & 0
\end{array}\right),
\end{eqnarray}
and $I_{2}$ and $I_{4}$ are the unit matrices of sizes 2 and 4, respectively.

%\begin{figure}[tbp]
%\begin{center}
%\includegraphics[scale=0.5]{8band.eps}
%\end{center}
%\caption{Schematic plot of the band structure near the fundamental gap for the eight-band Kane model.  }\label{setup}
%\end{figure}

The  matrix $H'_{8 \times 8}$ is then reduced to an effective Hamiltonian that depends only on the conduction band electron states
using a method similar to that used for the transformation of the Dirac equation of a four-spinor wave function into the Pauli-Schr\"odinger equation\cite{Winkler2006}. 
The Schr\"odinger equation 
\begin{eqnarray}
E\Psi = H'_{8\times 8} \Psi
\end{eqnarray}
%\begin{eqnarray}
%E\Psi = H'_{8\times 8} \Psi,
%\end{eqnarray}
with $\Psi = (\psi_{6c}, \psi_{8v}, \psi_{7v})^{T}$ 
reads 
\begin{eqnarray}
&&(\tE-V')\psi_{6c}=\sqrt3 P\v{T}\cdot \k_{g} \psi_{8v} -\frac{P\vsigma \cdot \k_{g}}{\sqrt3} \psi_{7v} \label{6c}\\
&&\psi_{8v}=\frac{1}{E_{G}}\Big(1+\frac{\tE-V'}{E_{G}}  \Big)^{-1} \sqrt3 P \v{T}^{\dagger}\cdot \k_{g} \psi_{6c} \label{8v}\\
&&\psi_{7v}=\frac{-1}{E_{G}+\Delta_{0}}\Big(1+\frac{\tE-V'}{E_{G}+\Delta_{0}}  \Big)^{-1} \frac{P \vsigma \cdot \k_{g}}{\sqrt3} \psi_{6c}. \label{7v} 
\end{eqnarray}
Inserting Eqs. (\ref{8v}) and (\ref{7v}) into (\ref{6c}), 
%Eliminating $\psi_{7v}$ and $\psi_{8v}$ from the coupled equation,  
\begin{eqnarray}
& \Big[ &
\v{T}\cdot \k_{g} \frac{3 P^{2}}{E_{G}} \Big(1+\frac{\tE-V'}{E_{G}}  \Big)^{-1}   \v{T}^{\dagger}\cdot \k_{g} \nonumber \\
&&
+\vsigma \cdot \k_{g} \frac{P^{2}/3}{E_{G} +\Delta_{0}} \Big(1+\frac{\tE-V'}{E_{G}+\Delta_{0}}  \Big)^{-1} \vsigma \cdot \k_{g}
\Big]\psi_{6c} \nonumber \\
&& = (\tilde{E}-V') \psi_{6c} \label{eq6c}
\end{eqnarray}
where $\tE=E-E_{c}$ and $E_{G}=E_{c}-E_{v}$.

To ensure norm conservation, a renormalized two component wave function $\tpsi_{6c}$ is sought that satisfies 
\begin{eqnarray}
\int d^{3}x \, |\Psi|^{2} = \int d^{3}x \, | \tpsi_{6c} |^{2}.\label{norm}
\end{eqnarray}
From Eqs. (\ref{8v}) and (\ref{7v}), the l.h.s. of  Eq. (\ref{norm}) is
\begin{eqnarray}
\int d^{3}x \, |\Psi|^{2} \approx
&& \int d^{3}x \, 
\Big| 1 + \frac{3P^{2} (\v{T}\cdot \k_{g}) (\v{T}^{\dagger}\cdot \k_{g})}{E_{G}^{2}} \nonumber\\
 &&+ \frac{P^{2} (\vsigma\cdot \k_{g}) (\vsigma \cdot \k_{g})}{3(E_{G}+\Delta_{0})^{2}} \Big|  |\psi_{6c}|^{2}.
\end{eqnarray}
Here we neglect $(\tilde{E}-V')/E_{G}$ and $(\tilde{E}-V')/(E_{G} + \Delta_{0})$. 
Hence, we find
$\tilde{\psi}_{6c} = 
\Big[1+ N\Big] \psi_{6c} 
$ with
\begin{eqnarray}
N=
\frac{3P^{2} (\v{T}\cdot \k_{g}) (\v{T}^{\dagger}\cdot \k_{g})}{2E_{G}^{2}} 
+
\frac{P^{2} (\vsigma\cdot \k_{g}) (\vsigma \cdot \k_{g})}{6(E_{G}+\Delta_{0})^{2}}.
\end{eqnarray}
Note that $N$ is essential to construct the wave function of the conduction electron because we need to keep the unitarity of the full wave function. Otherwise, we violate the unitarity of the original wave function, and therefore, both the eigenstates and eigenvalues of the Hamiltonian for the conduction electron cannot be calculated properly.

Inserting $\psi_{6c} \approx \Big[1- N\Big] \tpsi_{6c}$ into Eq. (\ref{eq6c}), we obtain an equation of the conduction band:
\begin{eqnarray}
(H'_0 + \delta H') \tpsi_{6c} = \tE \tpsi_{6c}, 
\end{eqnarray}
where the bare Hamiltonian $H'_{0}$ is defined by 
\begin{eqnarray}
H'_{0} &=& \frac{\hbar^{2} \k_{g}^{2}}{2m}+ V- \hbar \vsigma \cdot \Omega/2   \nonumber\\
&& + q\lambda \vsigma \cdot (\k_{g} \times \E) + q\lambda \div \E/2,
\end{eqnarray}
and $\delta H'$ is given by
\begin{eqnarray}
\delta H' &=& \frac{P^{2}}{3}\Big( \frac{2}{E_{G}} + \frac{1}{E_{G}+\Delta_{0}} \Big) \k_{g}^{2}  \nonumber\\
&&-\frac{P^{2}}{3} \Big( \frac{1}{E_{G}} - \frac{1}{E_{G}+\Delta_{0}} \Big)\frac{e}{\hbar} i \vsigma \cdot  (\k_{g} \times \k_{g}) \nonumber\\
&& +\frac{eP^{2}}{3} \Big( \frac{1}{E_{G}^{2}} - \frac{1}{(E_{G}+\Delta_{0})^{2}}\Big) \vsigma \cdot (\k_{g} \times \E )  \nonumber\\
&& -\frac{eP^{2}}{6} \Big( \frac{2}{E_{G}^{2}} + \frac{1}{(E_{G}+\Delta_{0})^{2}}   \Big)\div \E,     \label{tH'}
\end{eqnarray}
with $\E=(-1/e)\nabla V$. 
Here, we use the following relations:
\begin{eqnarray}
&&(\vsigma \cdot \k_{g})(\vsigma \cdot \k_{g}) = \k_{g}^{2}  +i \vsigma \cdot (\k_{g} \times \k_{g}), \\
&& 9(\v{T} \cdot \k_{g})(\v{T}^{\dagger} \cdot \k_{g} ) = 2 \k_{g}^{2}  -i \vsigma \cdot (\k_{g} \times \k_{g}).
\end{eqnarray}

Thus, we obtain the total Hamiltonian for the conduction band: $H'^{*} = H'_{0} + \delta H'$ which reads 
\begin{eqnarray}
H'^{*} &=& \frac{\hbar^{2} \k_{g}^{2}}{2m^{*}} + V - (1+\delta g)\frac{\hbar}{2}\vsigma \cdot \vOmega \nonumber\\
 &+& q(\lambda +\delta \lambda_{\rm S})\vsigma \cdot ( \k_{g} \times \E) 
      + \frac{q}{2}(\lambda + \delta \lambda_{\rm D}) \div \E. \label{r-H}
\end{eqnarray}
Here, 
the effective mass $m^{*}$, $\delta g$, $\delta \lambda_{\rm S}$, and $\delta \lambda_{\rm D}$ are given by 
\begin{eqnarray}
\frac{1}{m^{*}} &=& \frac{1}{m} + \frac{2P^{2}}{3\hbar^{2}}\Big( \frac{2}{E_{G}} + \frac{1}{E_{G}+\Delta_{0}} \Big)\\
\delta g &=&  -\frac{4m}{\hbar^{2}} \frac{P^{2}}{3} \Big( \frac{1}{E_{G}} - \frac{1}{E_{G}+\Delta_{0}} \Big),\label{r-g}\\
\delta \lambda_{\rm S} &=&  - \frac{P^{2}}{3} \Big( \frac{1}{E_{G}^{2}} - \frac{1}{(E_{G}+\Delta_{0})^{2}}\Big),\label{r-so}\\
\delta \lambda_{D} &=& \frac{P^{2}}{3} \Big( \frac{2}{E_{G}^{2}} + \frac{1}{(E_{G}+\Delta_{0})^{2}} \Big).\label{r-da}
\end{eqnarray} 
These factors $\delta g$, $\delta \lambda_{\rm S}$, and $\delta \lambda_{\rm D}$ are the same as those in an inertial frame in the presence of the magnetic field\cite{Kane1957,Winkler2006,Roth1959}. 
We emphasize that the renormalization of the spin-rotation coupling can be calculated from the renormalized $g$ factor.  
According to the conventional $\k \cdot \p$ method, the renormalized Zeeman term is 
\begin{eqnarray}
\frac{(g_{0}+\delta g)}2 \mu_{B} \vsigma \cdot \B
\end{eqnarray}
where $g_{0}=2$ is the bare $g$ factor. 
The renormalized $g$ factors $g_{0}+\delta g$ of the zincblende-type semiconductors have already been studied theoretically and experimentally. Therefore, the renormalization of the spin-rotation coupling can be calculated by using the well-known values. 
%The difference comes from the fact that the commutator of the kinetic momentum operator due to the rotation $(ie/\hbar)\k_{g} \times \k_{g} = \vsigma $
%the $SU(2)$ scalar potential in $\tH'$, as shown in the second line of Eq. (\ref{tH'}), is not proportional to that in Eq. (\ref{H'su2u1});
%$i \frac{e}{\hbar} \vsigma  \cdot  (\k' \times \k' ) = \vsigma \cdot (\B + \B_{g}) \neq \vsigma \cdot (\B + \B_{g}/2)$.
%The difference comes from the fact that the orbital and spin angular momentum, $\v{L}$ and $\v{S}=\hbar \vsigma/2$, couple to the magnetic field as $(\v{L} + 2\v{S}) \cdot \B$ whereas to the mechanical rotation as $(\v{L} + \v{S}) \cdot \vOmega$. Let us recall that the gauge potential $\A_{g}$ in the kinetic momentum $\k_{g}$ originates from the coupling $\v{L} \cdot \vOmega$. The orbital angular momentum $\v{L}$ contributes to the renormalization of $\v{S}\cdot \vOmega$ through the off-diagonal matrix elements which are proportional to the Kane momentum $P$. This situation is expressed by the third term in Eq. (\ref{tH'}). Similarly, the renormalization of the Zeeman interaction is caused by the contribution of the orbital part $\v{L}\cdot \B$ to the spin part $2\v{S}\cdot \B$ and the factor $2$ yields the difference of the renormalization. 

The difference between the renormalization factor of spin-rotation coupling and that of Zeeman coupling 
can be more clearly derived from the Pauli-Schr\"odinger Hamiltonian in the rigidly accelerating frame with external electromagnetic fields\cite{Mamoru2011ab}:  
\begin{eqnarray}
H'=\frac{\left\{ \p -q\A' - (qm\lambda/\hbar) \vsigma \times \E' \right\}^{2}}{2m} + q A'_{0} + \mu_{B}\vsigma \cdot \B',\nonumber \\ \label{Hmm}
\end{eqnarray}
where
$\A' = \A +\A_{g}$, 
$A_{0}' = \phi_{g} + (1/q)V +  A_{0}  + \gamma_{0}^{-1} \va \cdot \r  + \frac{m \lambda}{2} \div \E'    - \frac{m q \lambda^{2}}{\hbar^{2}}\E'^{2}$, 
$\E' = \E + (\Omega \times \r)\times \B $, 
$\B' = \B + \B_{g}/2$, $A_{\mu} = (A_{0},\A)$ is the electromagnetic field, and $\v{a}$ is the linear acceleration. 
The first term indicates that 
the conventional kinetic momentum $\hbar\k= \p -q\A$ is replaced by $\hbar \k'=\p - q\A'$ in the accelerated frame.
Therefore, if we start with Eq. (\ref{Hmm}), the renormalized Hamiltonian $H'^{*}$ is modified as
\begin{eqnarray}
H'^{*} &=& \frac{\hbar^{2} \k'^{2}}{2m^{*}} + V' + \left( 1+\frac{\delta g}{2}\right) \mu_{B} \vsigma \cdot \B - (1+\delta g)\frac{\hbar}{2}\vsigma \cdot \vOmega \nonumber\\
 &+& q(\lambda +\delta \lambda_{\rm S})\vsigma \cdot ( \k' \times \E') 
      + \frac{q}{2}(\lambda + \delta \lambda_{\rm D}) \div \E'. \label{r-H2}
\end{eqnarray}
The difference between the Zeeman interaction and the spin-rotation coupling comes from the fact that 
the bare Zeeman interaction is $\mu_{B} \vsigma \cdot \B$ while the bare spin-rotation coupling is $\mu_{B} \vsigma \cdot \B_{g}/2$. 
Detailed discussion on the Hamiltonian $H'$ in terms of  the $SU(2) \times U(1)$ gauge theory is given in Appendix.

\section{Frequency shift due to rotation in ESR}
The third and fourth terms in Eq. (\ref{r-H2}) can be rewritten as:
\begin{eqnarray}
\left( 1+ \frac{\delta g}{2} \right) \mu_{B} \vsigma \cdot ( \B + \Delta \B) 
\end{eqnarray}
where an effective magnetic field $\Delta \B$ is defined by
$\Delta \B = - \frac{2 + 2 \delta g}{2 + \delta g} \frac{\vOmega}{\gamma_{0}}. \label{DeltaB}$
In Ref. \cite{Lendinez2010}, 
the frequency shift of ESR due to rotation was pointed out as $\omega'_{\rm ESR}=\omega_{\rm ESR} - \Omega$ for an isotropic crystal.
By taking account of the interband mixing, 
the shift reads
\begin{eqnarray}
\omega'_{\rm ESR}=\omega_{\rm ESR} - (1+\delta g)\Omega. \label{ESRshift}
\end{eqnarray}

\section{Spin precession in a rotating body}
Let us consider effects of mechanical rotation on magnetization. 
The SOI is neglected for simplicity.
From the commutation relationship, 
the equation for spin precession in a rotating frame can be derived
\begin{eqnarray}
\dot{\vsigma} 
%&=&\frac{1}{i\hbar} \left[ \vsigma,  \frac{g^{*}_{S}}{2} \mu_{B} \vsigma \cdot \B - \frac{g^{*}_{SR}}{2} \frac{\hbar}{2}\vsigma \cdot \vOmega   \right] \nonumber\\
&=& - \left(1+\frac{\delta g}{2}\right)\mu_{B}\vsigma \times \B 
+ (1+\delta g) \frac{\hbar}{2}\vsigma \times \vOmega. \label{SPr}
\end{eqnarray}
Because of the renormalized spin-rotation coupling, the torque term due to the mechanical rotation has the factor $1 +   \delta g$.

\section{Discussions}
In the derivation above, we first include effects of the gauge field due to the rotation into the Hamiltonian in vacuum, 
and then apply the $\k \cdot \p$ perturbation with the crystal momentum $\k' = \p -q\A'$ and perform the projection to the conduction electron. 
Because the presence of the gauge field alters the translational symmetry mentioned earlier, 
the order of the introduction of the gauge field and the projection is essential to construct an effective theory, retaining the fundamental symmetry of the original system. 
Instead, if one started with the $\k \cdot \p$ perturbed Hamiltonian, $H^{*}$, and performed the unitary transformation
$U=\exp[i \v{J} t/\hbar]$, 
one would obtain the Hamiltonian:
$H^{*}_{R} = U H^{*} U^{\dagger} -i\hbar  U \frac{\del U^{\dagger}}{\del t} = H^{*} -  \r \times \p \cdot \vOmega - \frac{\hbar}{2} \vsigma \cdot \vOmega$,
where the spin-rotation coupling would not be renormalized, namely, the factor $1 +   \delta g$ missing.
Technically, this is because rotational effects to the off-diagonal matrix elements between the conduction and valence bands are not taken into account properly. 

We must care about such an order whenever we treat an effective Hamiltonian which is reduced from the original Hamiltonian.
If one can use the full set of the vectors in the original Hilbert space, it is not necessary to care about the order. 
However, when we use a reduced Hamiltonian, the Hilbert space is smaller than the original one. Therefore, the order is essential to obtain the correct answer. 
Similar situations are given in the following examples:
\begin{enumerate}
\item Derivation of the Pauli equation from the Dirac equation.  

To obtain the Pauli equation which includes the Zeeman and spin-orbit interactions, firstly, we introduce the gauge potential by replacing the canonical momentum $\p$ for the kinetic momentum $\p-q\A$. After that, we reduce the Dirac equation to the Pauli equation using the low energy expansion. Otherwise, one obtains neither Zeeman term nor spin-orbit interaction.
\item Conventional Kane model.  

If one starts with the eight band Kane model with the canonical momentum $\k$, namely, the momentum without the gauge potential $\A$, one obtains the conduction band Hamiltonian without renormalized $g$ factor. 
\end{enumerate}

%In the presence of the magnetic field, if one started with the $\k \cdot \p$ perturbed Hamiltonian with the canonical momentum operator $\hbar \k = \p$, and then introduced the gauge field $\A$ by replacing $\p$ for $\p -q \A$, the $g$ factor would remain unrenormalized. 

%\emph{Electron $g$ factor in semiconductors.}--- 
Finally, we mention the enhancement factor of the spin-rotation coupling for semiconductors. 
The electron $g$ factor in semiconductors has been widely investigated for 
low-dimensional semiconductor nanostructures such as GaAs/Al$_{x}$Ga$_{1-x}$ and Ga$_{1-x}$In$_{x}$As/InP heteropairs\cite{Kiselev1998}, 
and semiconductor quantum dots\cite{Pryor2006}.
%In Ref. \onlinecite{XWZhangAPL2007}, a very large $g$ factor with an absolute value larger than 900 is predicted for InSb$_{1-s}$N$_{s}$ bulk material using the ten-band $\k \cdot \p$ model. 
For lightly doped n-InSb at low temperature, $g \approx -49$ has been employed in a recent experiment\cite{Jaworski2012}. 
In this case, $\delta g = -51$, and then, the rotational ESR shift due to the 100kHz rotor is estimated to be 5MHz from Eq. (\ref{ESRshift}). 
This enhancement will be observed in experiment with a ultra high-speed rotor\cite{Ono2009}. 

%Let us consider the detectability of the predicted enhancement of the spin-rotation coupling in the ESR.  The ESR measurements are done in the gigahertz range and the width of the resonance is about as low as a few megahertz in some cases. Meanwhile the position of the ESR maximum can be determined with an accuracy of several hundreds kilohertz\cite{Lendinez2010} which is accessible by the high speed rotor\cite{Ono2009}. 
%Moreover, thanks to the enhancement, the rotational Doppler shift due to the 100KHz rotor will be 5MHz in the n-InSb.    
%Therefore, we conclude that the enhancement of the coupling will be measured in the ESR. 	

\section{Conclusion}
We have investigated enhancement of the spin-rotation coupling due to the interbandmixing in zincblende-type semiconductors. 
The Bloch's theorem and the envelope function approximation in the presence of the mechanical rotation have been constructed with the generalized crystal momentum which includes the gauge potential due to mechanical rotation. 
The effective Hamiltonian for conduction electrons in an accelerated body was derived from the $8\times 8$ Kane Hamiltonian. 
The Zeeman, spin-rotation coupling, SOI and Darwin terms in the accelerated frame are renormalized by interband mixing.  
The renormalized spin-rotation coupling provides the enhancement of the frequency shift in ESR and the mechanical torque in spin precession due to the interband mixing, which will be observed in experiment. 

\acknowledgments
The authors thank K. Harii, M. Ono, H. Chudo and E. Saitoh for discussions on experimental aspects.  
We also acknowledge valuable discussions with J. Suzuki and T. Ziman.
This work was supported by a Grant-in-Aid for Scientific Research from MEXT, Japan.

\appendix

\section{SU(2) $\times$ U(1) gauge theory in accelerated frames}
Here we formulate a $SU(2) \times U(1)$ gauge theory in accelerated frames, which offers a unified description of inertial effects on charge and spin currents.
It is well-known that the Pauli-Schr\"odinger equation can be analyzed by the $SU(2) \times U(1)$ gauge theory where the $U(1)$ gauge potential originates from electromagnetic fields and the $SU(2)$ counterpart from the Zeeman and spin-orbit interactions\cite{Anandan1989,Frohlich1993,JinLiZhan2006,Hatano2007}.
The $SU(2) \times U(1)$ gauge theory offers clear description of charge and spin physics in a unified way.
 
\label{sec:1}
\subsection{Pauli-Schr\"odinger equation}
We start with 	
the Hamiltonian for an electron in a rigidly accelerated frame in the presence of external electromagnetic fields\cite{Mamoru2011ab}:
\begin{eqnarray}
H'&=& \frac{(\p-q\v{A})^{2}}{2m}  + qA_{0} +m\va \cdot \r - \vb{\Omega}\cdot \r \times (\p-q\v{A}) \nonumber \\
&+& \mu_{B} \vb{\sigma} \cdot \v{B} - \frac{\hbar}{2} \vb{\sigma} \cdot \vb{\Omega} \nonumber \\
&+&\frac{q \lambda}{2 \hbar} \vb{\sigma}\cdot \Big[ (\p-q\v{A})\times \v{E}' - \v{E}' \times (\p-q\v{A})\Big] \nonumber \\
&+&\frac{q \lambda}{2}  \div \v{E}',  \label{H}
\end{eqnarray}
where
\begin{eqnarray}
\v{E}'=\v{E} +  \gamma_{0}^{-1} \va+ (\vOmega \times \r) \times \B.
\end{eqnarray}
Here, prime indicates the values including inertial effects.
The electron charge-to-mass quotient $\gamma_{0} = q/m$ is defined by the bare mass of the electron $m$ and  the charge $q=-e$, 
$\hbar$ the Planck constant divided by $2\pi$, $\mu_{B}=q\hbar/2m$ the Bohr magneton, $(A_{0},\v{A})$ the U(1) gauge potential, $\v{E}$ and $\v{B}$ the electric and magnetic fields, $\va$ and $\vOmega$ the linear acceleration and rotation frequency, $\vsigma$ the Pauli matrix, and $\lambda=\hbar^{2}/4m^{2}c^{2}$ the bare spin-orbit coupling.

The third and fourth terms in Eq. (\ref{H}) are inertial potentials due to linear acceleration and rotation respectively. Especially, the latter reproduces the Coriolis, centrifugal and Euler forces. 
The fifth term is the Zeeman term and the sixth term the spin-rotation coupling. These terms can be combined as
\begin{eqnarray}
 \mu_{B} \vb{\sigma} \cdot \v{B} - \frac{\hbar}{2} \vb{\sigma} \cdot \vb{\Omega}=\mu_{B} \vb{\sigma} \cdot \v{B}' \label{B'}
\end{eqnarray}
where 
\begin{eqnarray}
\v{B}' &=& \v{B} + \B_{S},
\end{eqnarray}
and 
\begin{eqnarray}
\B_{S} &=&\frac{\vOmega}{\gamma_{0}}.
\end{eqnarray}
Thus, the spin-rotation coupling can be interpreted as the Zeeman interaction due to an effective magnetic field $\B_{S}$, 
which is often referred as the Barnett magnetic field\cite{Barnett1915}.
The last two terms in Eq. (\ref{H}) are the spin-orbit interaction and the Darwin term augmented by the inertial effects. 
The spin-orbit interaction is responsible for spin current generation by mechanical means\cite{Mamoru2011ab}.

\subsection{$SU(2)\times U(1)$ gauge theory}
It is well-known that Hamiltonian with spin-orbit interaction can be rewritten as the $SU(2) \times U(1)$ gauge theory which allows us to understand physics of both charge and spin current in a unified way\cite{Anandan1989,Frohlich1993,JinLiZhan2006,Hatano2007}. 
The Hamiltonian (\ref{H}) can be written as:
\begin{eqnarray}
H'=\frac{1}{2m}(p_{i} -qA'_{i}-\eta \CA'^{a}_{i} \tau^{a})^{2} + q A'_{0} + \eta \CA'^{a}_{0}\tau^{a},
\end{eqnarray}
where
\begin{eqnarray}
A'_{i} &=& A_{i} -\gamma_{0}^{-1}  \epsilon_{ijk }\Omega_{j} \times r_{k}, \\
A_{0}' &=& A_{0} + \gamma_{0}^{-1} \va \cdot \r  + \frac{m \lambda}{2} \div \E'  \nonumber\\
&&-\frac{m}{2}(\vOmega \times \r)^{2}   - \frac{m q \lambda^{2}}{\hbar^{2}}\E'^{2}, \\
\CA'^{a}_{i} &=& - \frac{2 mq\lambda }{\eta \hbar} \epsilon_{aji} E'_{j}, \\
\CA'^{a}_{0} &=& \frac{2\mu_{B}}{\eta} B'_{a}.
\end{eqnarray}
Here $\tau^{a}= \sigma_{a}/2$ is the generator of the $SU(2)$ Lie group and we choose the ``$SU(2)$ charge'' as $\eta = \hbar$.
This Hamiltonian consists of two gauge potentials; 
the $U(1)$ gauge potential denoted by $A'_{\mu}=(A'_{0},A'_{i}) (i=1,2,3)$ 
and the $SU(2)$ potential by $\CA'^{a}_{\mu}=(\CA'^{a}_{0},\CA'^{a}_{i})$.
These relations shows that 
all of the inertial effects on spin and charge due to the linear acceleration $\va$ and the rotation $\vOmega$ can be expressed by the $SU(2)$ and $U(1)$ gauge potentials.

\subsubsection{Gravitomagnetic field vs Barnett magnetic field}
%According to the Einstein's equivalence principle, gravitation cannot be distinguished locally from inertial effects, and weak gravity

The conventional magnetic field $\B=\nabla \times \A$ is modified by the rotation $\vOmega$ in the $U(1)$ gauge field as
\begin{eqnarray}
\nabla \times \A' = \B + \B_{g},
\end{eqnarray}
with
\begin{eqnarray}
\B_{g}= 2\gamma_{0}^{-1}\vOmega.
\end{eqnarray}
The effective magnetic field $\B_{g}$ is called the gravitomagnetic field in the context of general relativity\cite{RyderGR2009}.
On the contrary, 
$\B$ in the time component  of $SU(2)$ gauge potential $\CA_{0}$, which originates from the Zeeman term,
is modified by the Barnett magnetic field $\B_{S}=\gamma_{0}^{-1}\vOmega$ as shown in Eq. (\ref{B'}). 
Note that the inertial effect due to the rotation on charge is different from that on spin by the factor 2.
To understand this difference, 
it is useful to give another expression of 
the gravitomagnetic and Barnett fields by introducing two kinds of $g$ factors defined by
\begin{eqnarray}
\begin{cases}
g_{L}  = 1 & (\mbox{orbital angular momentum})\\
g_{S}  = 2 & (\mbox{spin angular momentum}).
\end{cases}
\end{eqnarray}
and gyromagnetic ratios by
\begin{eqnarray}
\gamma_{L}  &=& \frac{g_{L} \mu_{B}}{\hbar}, \\
\gamma_{S}  &=& \frac{g_{S} \mu_{B}}{\hbar} 
\end{eqnarray}

Then, $\B_{g}$ and $\B_{S}$ can be written as
\begin{eqnarray}
\B_{g} &=& - \frac{\vOmega}{\gamma_{L}},  \\
\B_{S} &=& - \frac{\vOmega}{\gamma_{S}}.
\end{eqnarray}
Using these expressions, generator of the mechanical rotation, $(\v{L} + \v{S}) \cdot \vOmega$, is rewritten as
\begin{eqnarray}
(\v{L} + \v{S}) \cdot \vOmega &=& \v{m}_{L} \cdot \B_{g} + \v{m}_{S} \cdot \B_{S},
\end{eqnarray} 
where $\v{L}$ and $\v{S}$ are the orbital and spin angular momentum and magnetic moments are given by
\begin{eqnarray}
\v{m}_{L} &=& - \gamma_{L} \v{L}, \\
\v{m}_{S} &=& - \gamma_{S} \v{S}. 
\end{eqnarray}

\subsubsection{Anomalous velocity}
The spatial component of the $SU(2)$ gauge potentials $\CA'^{a}_{i}$ yields 
spin-dependent velocity (anomalous velocity) due to inertial effects\cite{Mamoru2011ab}
\begin{eqnarray}
v'_{i} &=& \frac{1}{i\hbar} [r_{i},H] \nonumber\\
&=& v_{c,i} + v_{s,i} 
\end{eqnarray}
where the spin-independent velocity $v_{c,i}$ and the spin-dependent velocity $v_{s,i}$ are given by
\begin{eqnarray}
v_{c,i}&=&\frac{ p_{i} -qA'_{i}}{m}  \nonumber\\
v_{s,i}&=&\frac{q\lambda}{\hbar} \epsilon_{ijk} \sigma_{j} E'_{k}.
\end{eqnarray}
In the case that $A_{\mu} =0$, $\vOmega = \v{0}$, and $\va  \neq \v{0}$,
the anomalous velocity becomes
\begin{eqnarray}
\vv_{s} = \frac{m \lambda}{\hbar} \vsigma \times \va.
\end{eqnarray} 
Then, spin current is generated perpendicular to the effective electric field due to the linear acceleration $\gamma^{-1}_{0} \va$.
In the case that $\B \neq \v{0}$, $\vOmega \neq \v{0}$, and $\B \parallel \vOmega$, spin current is generated perpendicular to the effective electric field $(\B\cdot \vOmega) \r$ since the anomalous velocity reads
\begin{eqnarray}
\vv_{s} = \frac{q\lambda}{\hbar} \vsigma \times (\B\cdot \vOmega) \r.
\end{eqnarray}

\subsection{Lagrangian formalism and Noether current}
According to the Noether's theorem, continuous symmetries of a system are closely related to conserved quantities. 
Noether's theorem in the $SU(2) \times U(1)$ gauge theory gives the relation between charge and charge current as well as spin and spin current. However, unlike to charge current, spin current is not conserved quantity because of the non-Abelian properties of the $SU(2)$ gauge group.

The Lagrangian density for the $SU(2) \times U(1)$ gauge theory in accelerated frames is given by
\begin{eqnarray}
\CL'=&&\frac{ i }{ 2 }(\dot{\psi}^{\dagger} \psi-\psi^{\dagger} \dot{\psi}) + \psi^{\dagger} (qA'_{0}+\eta \CA'^{a}_{0}\tau^{a})\psi \nonumber\\
&& + \frac{m}{2} (v'_{i} \psi)^{\dagger} (v'_{i} \psi) \nonumber \\
&& -\frac14 F'_{\mu\nu} F'_{\mu\nu} -\frac14 \CF'^{a}_{\mu\nu} \CF'^{a}_{\mu\nu},  
\end{eqnarray}
where
\begin{eqnarray}
v'_{i} &=& \frac{1}{m}(p_{i} -qA'_{i}-\eta \CA'^{a}_{i} \tau^{a}),\\
F'_{\mu\nu} &=& \del_{\mu}A'_{\nu} -\del_{\nu} A'_{\mu},\\
\CF'^{a}_{\mu\nu} &=& \del_{\mu}\CA'^{a}_{\nu} -\del_{\nu} \CA'^{a}_{\mu} + \eta \epsilon^{abc} \CA'^{b}_{\mu} \CA'^{c}_{\nu}.
\end{eqnarray}
The Noether currents associated with the $U(1)$ and $SU(2)$ gauge symmetry are given by
\begin{eqnarray}
J'_{\mu} &=& \frac{\del \CL'}{\del A^{\mu}} \nonumber \\
&=& e(\psi^{\dagger} \psi,{\rm Re}[\psi^{\dagger} v'_{i} \psi])
\end{eqnarray}
and
\begin{eqnarray}
\CJ'^{a}_{\mu} &=& \frac{\del \CL'}{\del \CA'^{a\mu}} \nonumber \\
&=&\eta 
\left(
\psi^{\dagger} \tau^{a} \psi , {\rm Re} \left[ \psi^{\dagger} \frac{\tau^{a}v'_{i} + v'_{i} \tau^{a}}{2}  \psi \right]  
\right) \nonumber \\
&& + \eta \epsilon^{abc} \CA'^{b}_{\nu} \CF'^{c}_{\mu\nu}.
\end{eqnarray}
Then, we obtain the continuity relation in accelerated frames for charge 
\begin{eqnarray}
\del^{\mu}J'_{\mu}=0 \label{charge-c}
\end{eqnarray}
and for spin 
\begin{eqnarray}
\del^{\mu}\CJ'^{a}_{\mu}=0.\label{spin-c}
\end{eqnarray}
The relation (\ref{charge-c}) reads
\begin{eqnarray}
\frac{\del \rho'_{c}}{\del t} + \div \v{j}'_{c} =0 
\end{eqnarray}
where 
\begin{eqnarray}
j'_{c,\mu}&=&(\rho'_{c},j'_{c,i}) \nonumber\\
&=&  e(\psi^{\dagger} \psi,{\rm Re}[\psi^{\dagger} v'_{i} \psi]).
\end{eqnarray}
On the contrary, Eq. (\ref{spin-c}) reads
\begin{eqnarray}
\frac{\del \rho'_{s}}{\del t} + \div \v{j}'_{s} =  - \del^{\mu} (\eta \epsilon^{abc} \CA'^{b}_{\nu} \CF'^{c}_{\mu\nu} ) \neq 0 \label{spin-c2}
\end{eqnarray}
where 
\begin{eqnarray}
j'^{a}_{s,\mu} &=& (\rho'^{a}_{s},j'^{a}_{s,i}) \nonumber\\
&=& \eta 
\left(
\psi^{\dagger} \tau^{a} \psi , {\rm Re} \left[ \psi^{\dagger} \frac{\tau^{a}v'_{i} + v'_{i} \tau^{a}}{2}  \psi \right]  
\right).
\end{eqnarray}
Thus, the spin current in accelerated frames $\v{j}'_{s}$ is not conserved quantity because of the r.h.s of Eq. (\ref{spin-c2}) which originates from the non-Abelian properties of the $SU(2)$ gauge potential $\CA'^{a}_{\mu}$.

\subsection{Inertial forces acting on charge and spin current}
We can identify the inertial forces acting on charge and spin current 
on the basis of the analogy with the Lorentz force in the conventional electromagnetism:
\begin{eqnarray}
f_{c,i} = j_{c, \mu} F_{\mu i},
\end{eqnarray} 
where $j_{c,\mu}$ is the 4-vector of the charge current and the field strength of the conventional electromagnetic field $F_{\mu \nu} = \del_{\mu} A_{\nu} - \del_{\nu} A_{\mu}$.
In the $SU(2) \times U(1)$ gauge theory in accelerated frames, we have 
the inertial force acting on charge and charge current:
\begin{eqnarray}
f'_{c,i} &=& j'_{c,\mu} F'_{\mu i},\label{f'-c}
\end{eqnarray}
and that on spin and spin current:
\begin{eqnarray}
f'_{s,i} &=& j'^{a}_{s,\mu} \CF'^{a}_{\mu i} \nonumber\\
 &=& \rho'^{a}_{s} \CE'^{a}_{i} +  \epsilon_{ijk}j'^{a}_{s,j} \CB'^{a}_{k},\label{f'-s}
\end{eqnarray}
where the $SU(2)$ ``electric'' field $\CE'^{a}_{i}$ and ``magnetic'' field $\CB'^{a}_{k}$ are given by
\begin{eqnarray}
\CE'^{a}_{i} &=& \del_{0}\CA'^{a}_{i}   -\del_{i}\CA'^{a}_{0}, \\
\CB'^{a}_{k} &=& \epsilon_{klm} \del_l \CA'^{a}_{m}.
\end{eqnarray}
Equation (\ref{f'-c}) is a unified expression of the Lorentz force and the inertial force including the Coriolis force with the relativistic correction due to the Darwin term. Equation (\ref{f'-s}) is also a unified expression of spin-dependent inertial force\cite{Mamoru2011ab}.
%It is noted that the $SU(2)$ electric field $\CE'^{a}_{i}$ contains a mechanical analogue of the Stern-Gerlach force:
%\begin{eqnarray}
% - \rho'^{a}_{s} \del_{i}\CA'^{a}_{0} &=& - \rho'^{a}_{s} \del_{i} \left[ \frac{2\mu_{B}}{\eta}B'_{a} \right] \nonumber\\
% &=& \sigma_{a} \del_{i} \left[ 2\mu_{B} (B_{a} 
%         + \gamma_{0}^{-1}\Omega_{a} ) \right].\label{msg}
%\end{eqnarray}
%The second term in Eq. (\ref{msg}) is closely related to the mechanical generation of spin current in elastic bodies\cite{Mamoru2013SAW}.


\end{document}